\documentclass[12pt]{article}

\usepackage{epsfig}
\usepackage{amssymb}
\usepackage{amsmath}                          
\usepackage{graphics,graphpap}
\usepackage{graphicx}
\usepackage{epstopdf}

\usepackage{array} 
\usepackage{amssymb}
\usepackage{graphics,graphpap}
\usepackage{graphicx}
\usepackage{color}
\usepackage{dcolumn}
\usepackage{epsfig}
\usepackage{bm}
\usepackage{amsmath}
\usepackage{amsfonts}
\renewcommand{\thefootnote}{\fnsymbol{footnote}}

\newcommand{\mudmu}{\mu\frac{d}{d\mu}}  
\newcommand{\beq}{\begin{eqnarray}}
\newcommand{\eeq}{\end{eqnarray}}

\newcommand{\bmp}{\noindent\begin{minipage}{16cm}}
\newcommand{\emp}{\end{minipage}\vskip 7mm} 

\def\drawbox#1#2{\hrule height#2pt
        \hbox{\vrule width#2pt height#1pt \kern#1pt
              \vrule width#2pt}
              \hrule height#2pt}

\def\Asym#1#2{\vcenter{\vbox{\drawbox{#1}{#2}
              \kern-#2pt 
              \drawbox{#1}{#2}}}}

\newcommand{\tchiral}{\langle \bar q q \rangle}
\newcommand{\qbarq}{{\bar q q}}
\newcommand{\quark}{\langle \bar q q\rangle}

\newcommand{\Cdot}{\! \cdot \!}

\def\simge{\mathrel{%
   \rlap{\raise 0.511ex \hbox{$>$}}{\lower 0.511ex \hbox{$\sim$}}}}

\def\simle{\mathrel{
   \rlap{\raise 0.511ex \hbox{$<$}}{\lower 0.511ex \hbox{$\sim$}}}}

\def\s#1{\setbox0=\hbox{$#1$}%
\rlap{\ifdim\wd0>.7em\kern.22\wd0\else\kern.1\wd0\fi /}#1}

\newcommand{\matel}[3]{\langle #1|#2|#3\rangle}
\newcommand{\vev}[1]{\langle #1 \rangle}

\newcommand{\cU}{{\cal U}}

\newcommand{\lc}{{\Lambda_\mathrm{IR}}}
\newcommand{\refeq}[1]{Eq.~(\ref{eq:#1})}

\begin{document}

\begin{titlepage}
\begin{flushright}
\begin{tabular}{l}
  SHEP 10-02 \\
  CP3-Origins-2010-02
\end{tabular}
\end{flushright}

\vskip1.5cm
\begin{center}
  {\Large \bf \boldmath  Hyperscaling  relations in  \\
  mass-deformed conformal gauge theories} 
  \vskip1.3cm 
  {\sc Luigi Del Debbio$^{\,a}$\footnote{luigi.del.debbio@ed.ac.uk}  \&
    Roman Zwicky$^{\,b}$\footnote{Roman.Zwicky@soton.ac.uk}}
  \vskip0.5cm
  
  $^a$ {\sl School of Physics and Astronomy, University of Edinburgh, 
    Edinburgh EH9 3JZ, Scotland} \\
  $^b$ {\sl School of Physics \& Astronomy, University of Southampton, 
    Highfield, Southampton SO17 1BJ, UK} \, 
  \vspace*{1.5mm}
\end{center}

\vskip0.6cm

\begin{abstract}
  We present a number of analytical results which should guide the
  interpretation of lattice data in theories with an infra-red fixed
  point (IRFP) deformed by a mass term $\delta {\cal L} = - m \bar
  qq$.  From renormalization group (RG) arguments we obtain the
  leading scaling exponent, $F \sim m^{\eta_F}$, for all decay
  constants of the lowest lying states other than the ones affected by
  the chiral anomaly and the tensor ones. These scaling relations
  provide a clear cut way to distinguish a theory with an IRFP from a
  confining theory with heavy fermions.  Moreover, we present a
  derivation relating the scaling of $\tchiral \sim m^{\eta_{\bar
      qq}}$ to the scaling of the density of eigenvalues of the
  massless Dirac operator $\rho(\lambda) \sim \lambda^{\eta_{\bar
      qq}}$.  RG arguments yield $\eta_{\bar qq} =
  (3-\gamma_*)/(1+\gamma_*)$ as a function of the mass anomalous
  dimension $\gamma_*$ at the IRFP.  The arguments can be generalized
  to other condensates such as $\vev{G^2} \sim m^{4/(1+\gamma_*)}$.
  We describe a heuristic derivation of the result on the condensates,
  which provides interesting connections between different approaches.
  Our results are compared with existing data from
  numerical studies of SU(2) with two adjoint Dirac fermions.
\end{abstract}

\nonumber

\end{titlepage}

\newpage

\setcounter{footnote}{0}
\renewcommand{\thefootnote}{\arabic{footnote}}
\newcommand{\ym}{y_m}

\section{Introduction}
\label{sec:intro}
There are numerous examples of two-dimensional field theories that are
invariant under the full conformal group. In four dimensions, the beta
function of $\mathcal N=4$ super Yang-Mills is known to vanish to all
orders in perturbation theory, for any value of the coupling, so that
the theory is scale invariant.  Other theories have isolated zeroes of
the beta function that correspond to fixed points of the
renormalization group (RG) flow. For instance, the gauge coupling $g$
in QCD flows to zero as the energy scale is increased, leading to the
well-known phenomenon of asymptotic freedom; in this case $g=0$ is
commonly called an UV fixed point. On the other hand, if a theory has
an IR fixed point (IRFP), the couplings will flow to such a fixed
point at large distances, and the theory becomes scale invariant in
the large--distance regime. Theories with an IRFP do not break chiral
symmetry spontaneously, and are said to lie in the {\em conformal
window}.

Supersymmetric examples of theories within the conformal window have
been studied in detail - see e.g. Ref.~\cite{Intriligator:1995au} for
a review. Recently there has been a lot of interest in identifying
non-supersymmetric gauge theories with an IRFP.  The main motivation,
besides intrinsic interest, comes from the fact that theories {\em
near} the conformal window correspond to the class of theories
underlying walking
technicolor~\cite{Holdom:1983kw,Yamawaki:1985zg,Akiba:1985rr,Sannino:2009za,Luty:2004ye},
which is the phenomenologically most viable offspring of technicolor
theories~\cite{Weinberg:1975gm,Susskind:1978ms,Eichten:1979ah,Hill:2002ap}.
A chirally broken theory near the edge of the conformal window is
supposedly identified by an enhancement of the ratio
$\tchiral/f_\pi^3$ with respect to a QCD-like
theory~\cite{Hill:2002ap}.  Unfortunately this quantity does not
display a simple known parametric behaviour.  Another strategy, which
is adopted in this paper, is to first identify theories within the
conformal window, and then approach the boundary of the window using
the available information on the color-flavor phase
diagram~\cite{Holdom:1983kw,Yamawaki:1985zg,Luty:2004ye,Dietrich:2006cm}.
 
The identification of conformal theories using numerical simulations
is a difficult task, since the only observable quantities would be the
power-law scaling of correlators at large distances.  However actual
lattice simulations are performed in a finite volume, and with a
non-vanishing fermion mass; both the mass and the finite size of the
system are relevant operators at large distances and drive the theory
away from conformal behaviour.  Turning a technical limitation into a
tool, it has become a standard strategy to consider conformal gauge
theories (CGT) candidates deformed by a mass term, and to identify
them from the study of their hadronic observables.
Thus, if there exists an IRFP, the lattice results should be described
by a mass-deformed conformal gauge theory (mCGT), obtained by adding a
bare mass to the original lagrangian
\begin{equation}
\label{eq:mqq}
\delta {\cal L} =   - m \bar q q  \, . 
\end{equation}
As a consequence of the deformation, these theories are expected to
develop a mass gap and a fermion condensate and thus give rise to
asymptotic states and related observables, which scale to zero as the
massless limit is approached. For any observable ${\cal O}$ the
leading exponent $\eta_{\cal O}$ of the mass deformation is defined
from its scaling as $m\to 0$:
\begin{equation} 
\label{eq:Om}
{\cal O} \sim m^{\eta_{\cal O}} + \text{higher order
    in $m$} + \text{terms analytic in $m$}\, .
\end{equation}
These critical exponents can be measured on the lattice and it is the
aim of this work to provide predictions for them that can be tested
numerically. 

The paper is organized as follows. In section~\ref{sec:cw} we set the
framework by discussing some characteristics of theories inside the
conformal window. In section~\ref{sec:IRFP} we discuss general aspects
of IRFPs, and introduce the standard tools for analyzing the behaviour
of field correlators near a fixed point of the RG flow. Thereby we
obtain the hyperscaling relations that are usually derived in the
context of critical phenomena \cite{Cardy:1996xt}, and we study the information that they
yield in the framework of mCGT.

Section~\ref{sec:BCS} is devoted to the study of the chiral condensate
in mCGTs. First we review the relation between the scaling of the
chiral condensate with the fermion mass, and the density of
eigenvalues of the massless Dirac operator in the infinite--volume
limit.  As stated above, the chiral condensate must vanish as the
fermion mass is taken to zero at a rate that is dictated by a critical
exponent $\eta_\qbarq$. The non-analytic dependence of the fermion
condensate on the fermion mass is directly related to the scaling
exponent for the eigenvalue density of the massless Dirac operator. As
pointed out in Ref.~\cite{DeGrand:2009et}, the exponents turn out to
be the same:
\begin{equation}
  \label{eq:BC2}
  \quark \sim m^{\eta_{\qbarq}}  \quad \Rightarrow \quad \rho(\lambda) \sim
  \lambda^{\eta_{\qbarq}}\, .
\end{equation}
The scaling exponent $\eta_{\qbarq}$ is determined as a
function of $\gamma_*$, the anomalous dimension of the mass at the
IRFP. The RG analysis, which applies to all condensates, yields 
\begin{equation}
  \label{eq:BC3}
  \eta_{\qbarq}=\frac{(3-\gamma_*)}{(1+\gamma_*)}
\end{equation}
We then present
the determination of this coefficient from a heuristic calculation,
which provides some physical insight in the dynamics of mCGT. The
limitations of such a heuristic approach are highlighted, and the
interpretation of IR and UV cutoffs is clarified. We conclude this
section by analyzing current lattice data for the eigenvalue
distribution in an SU(2) gauge theory with two flavours in the adjoint
representation.

In section~\ref{sec:fPS} we explore the consequences of hyperscaling
for the decay constants of the hadronic states. Our results,
summarized in Tab.~\ref{tab:tab_decay}, can schematically written as,
\begin{equation}
G \sim m^{\frac{\Delta_{\cal O}-1}{1+\gamma_*}} \;, \qquad \matel{0}{{\cal O}(0)}{H(p)} = G \;,
\end{equation}
for operators with scaling dimension $\Delta$.
Further informations
are obtained by combining these results with the chiral Ward
identities in section~\ref{sec:WI}; these scaling predictions for the
decay constants are then compared with recent results from numerical
simulations of potential mCGT on the lattice. Finally we discuss the
implication of the scaling of the decay constants for the width of the
hadronic states, and compare the scaling of the decay constants in a
mCGT to the one of heavy quarkonia states in a chirally broken theory
like QCD.

\subsection{Conformal window - discussion and results}
\label{sec:cw}

It is well known that $\mathrm{SU}(N)$ gauge theories with $n_f$
fermions are asymptotically free as long as $n_f$ does not exceed an
upper limit that depends on the number of colours $N_c$ and the fermion
representation $R$. At small distances the gauge coupling decreases
logarithmically, and the dynamics is successfully described by
perturbation theory. In the $\mathrm{SU}(3)$ gauge theory minimally
coupled to $n_f=2$ light flavors in the fundamental representation,
the coupling increases at large distances, and the theory undergoes
confinement and spontaneous chiral symmetry breaking, exhibiting a
spectrum of bound states. In the massless limit, the spectrum includes
three massless Goldstone bosons, known as ($\pi^0, \pi^+,\pi^-$),
reflecting the spontaneous breaking of chiral symmetry. As a
consequence, there is a gap in the spectrum between the pions and the
rest of the states whose masses are parametrically of the order of
some hadronic scale $\Lambda \simeq
\Lambda_{\rm QCD}$, and remain finite in the chiral limit. At low
energies compared to $\Lambda$ the dynamics are successfully described
by an effective theory of self-interacting pions, known as chiral
perturbation theory. A small non-vanishing mass can easily be
incorporated as a perturbation of the massless theory.

As the number of light fermions is increased, before asymptotic
freedom is lost, the theory may develop an infrared fixed point (IRFP)
due to the effect of the fermions on the running of the coupling. We
shall denote by $n_{f,c}$ the number of fermions above which the
theory exhibits an IRFP. In this case the theory becomes
scale-invariant at large distances, while the short-distance behaviour
is still the one dictated by asymptotic freedom. As a consequence of
the scale invariance at large distances, the theory cannot be in a
confining phase and chiral symmetry remains unbroken. The
long-distance dynamics is governed by the critical exponents of the
IRFP, which determine the scaling laws in the vicinity of the fixed
point. The Banks-Zaks theories~\cite{Banks:1981nn}, where $N_c$ and
$n_f$ are arranged such that the critical coupling $g^* \ll 1$,
provide one working example of a theory within the conformal
window. Early studies of near conformal and IRFP theories were based
on approximate solutions of the Schwinger--Dyson
equations~\cite{Cohen:1988sq,Appelquist:1998xf}; these analyses were
extended to higher representations in Ref.~\cite{Dietrich:2006cm}.
Unfortunately it is very difficult to control the systematic errors
due to the truncation of the 1PI vertices appearing in the
Schwinger--Dyson equations. Moreover Schwinger--Dyson equations
predict the anomalous dimension of the mass to be around one, whereas
unitarity constraints on the conformal group~\cite{Mack:1975je}, in
principle, allow for $\gamma_* \leq 2$.

Recent results have appeared recently, that address this problem
either from an RG point of
view~\cite{Braun:2006jd,Braun:2009ns,Gies:2005as}, or from a
gauge/string duality
perspective~\cite{Nunez:2008wi,Elander:2009pk,Piai:2010ma}. We defer
the investigation of the connections between our results and these
other approaches for further studies.

Recent numerical simulations of gauge theories on the lattice have
triggered a renewed interest in those theories and in turn in
technicolor models.  Algorithmic progresses have made lattice
simulations with light dynamical fermions accessible on current
hardware~\cite{Hasenbusch:2002ai,Luscher:2005rx,Clark:2006fx}. This
opens the possibility to obtain first principles results for
technicolor, and several preliminary investigations have
appeared~\cite{Catterall:2007yx,Appelquist:2007hu,DelDebbio:2008wb,Shamir:2008pb,Deuzeman:2008sc,DelDebbio:2008zf,Catterall:2008qk,DeGrand:2008kx,Hietanen:2008mr,Appelquist:2009ty,Hietanen:2009az,Deuzeman:2009mh,Fodor:2009nh,DeGrand:2009mt,DeGrand:2009et,Hasenfratz:2009ea,DelDebbio:2009fd,Fodor:2009wk,Pica:2009hc,Fodor:2009ar}. It
is important to bare in mind that recent lattice results for theories
that may lie inside the conformal window are plagued by systematic
errors, and their interpretation still needs to be clarified. A recent
discussion of the lattice artefacts in simulations of theories with a
potential IRFP can be found in
Refs.~\cite{Pica:2009hc,Bursa:2009we}. For these theories, unlike in
QCD, there are no experimental results to guide the lattice
simulations.

Therefore it is crucial to develop analytical results in order to
guide the lattice studies, and help in analyzing their outcome. A
wider range of analytical predictions, together with more extensive
simulations, will help in finding robust evidence for the existence of
IRFPs.

\section{Infrared fixed points}
\label{sec:IRFP}
\begin{figure}[ht]
\centerline{\includegraphics[width=3.0in]{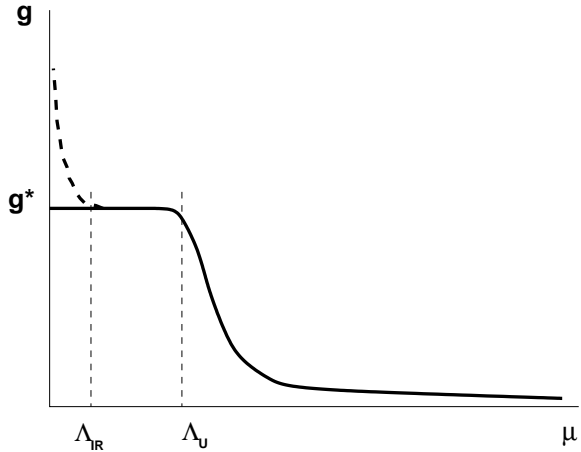}}
  \caption{\small Running of the coupling as a function of the energy scale
  for a theory with an IRFP. At low energies the coupling flows to a
  fixed-point value $g^*$, while the high energy behaviour is the
  usual one expected for asymptotically free theories. The scale
  $\Lambda_\cU$ corresponds to the energy where the running starts to
  be dictated by asymptotic freedom. The dashed curve at low energies
  shows the running of the coupling when a fermionic mass term is
  switched on. } \label{fig:betafn}
\end{figure}

Let us henceforth consider theories {\em inside} the conformal window,
{\it i.e.}  gauge theories minimally coupled to a number $n_f$ of
Dirac fermions, with the number of flavors and their representation
adjusted so that the theories are scale-invariant at large distances
when the fermions are massless. In general, fixed points of RG flows
are identified by the zeroes of the $\beta$ functions that describe
the evolution of dimensionless couplings.  The typical evolution of a
running coupling is sketched in Fig.~\ref{fig:betafn}. The running
coupling flows to a constant value at small energies, which
corresponds to a zero of the beta function. The value $g^*$ of the
coupling at the fixed point, and the precise shape of the
nonperturbative function $g(\mu)$ are scheme-dependent. However the
existence of the fixed point and the critical exponents are universal.

The fermion mass is a relevant coupling at the IRFP, and drives the
theory away from it. In a theory with a non-vanishing fermion mass,
the fermionic degrees of freedom decouple at low
energies, and the theory behaves like a pure Yang--Mills theory. The
running of the gauge coupling for the massive theory is given by the
dashed curve at small $\mu$ in Fig.~\ref{fig:betafn}, where the
running of the coupling below some scale $\lc$ is explicitly
drawn. Note that in the presence of an IRFP $\lc$ goes to zero as the
fermion mass vanishes.

The running of the mass is described by its anomalous dimension, which 
has the opposite sign of the anomalous dimension of the 
renormalized composite operator $\qbarq$,
\begin{equation}
  \label{eq:gammadef} \mudmu \left.\qbarq\right|_\mu = \gamma_{\bar q
  q}(\mu) \left.\qbarq\right|_\mu
  = \gamma(\mu) \left.\qbarq\right|_\mu \, .
\end{equation}
We have explicitly indicated the scale dependence of the various
quantities. In this paper we will use the symbol $\gamma$ to denote the
anomalous dimension of the mass and quark 
condensate: $\gamma \equiv \gamma_m = - \gamma_{\bar q q}$.

Note that the anomalous dimension away from the fixed point depends on
the renormalization scheme. However its value $\gamma_*$ at the IRFP
is a scheme-independent quantity. A concise discussion of the
scheme-dependent features of IRFPs can be found in
Ref.~\cite{Bursa:2009we}.  

Throughout this paper we will often refer to scaling dimensions of
operators,  denoted by $\Delta$; thery are obtained as
the sum of the naive mass dimension of the operator and the anomalous
dimension.  For example for the operator $\bar q q$ we write:
\begin{equation}
  \label{eq:Deltaym}
  \Delta_{\bar q q} = d_{\bar q q} + \gamma_{\bar q q} 
  = 3 - \gamma_* \;, \quad y_m = 1 + \gamma_*\, ,
\end{equation}
where we have also introduced the scaling exponent $y_m$, which often
appears in what follows and is widely used in the RG-literature 
\cite{Cardy:1996xt}. Throughout this paper we will use these
notations interchangeably.

Scaling laws are derived by assuming that the fermion mass is the only
relevant operator at the IRFP. RG equations will be used below in
order to derive the scaling of the chiral condensate as a function of
the fermion mass. It is therefore worthwhile to briefly recall how the
scaling relation for the masses in the spectrum is obtained. A recent
discussion of RG flows in the vicinity of an IRFP can be found in
Refs.~\cite{DeGrand:2009mt,DelDebbio:2010hu,DelDebbio:2010hx}.

Let us consider the zero-momentum vacuum correlator of an
interpolating field $H(x)$ with the quantum numbers of a given state
in the spectrum:
\begin{equation}
  \label{eq:twopt}
  C_{H}(t;g,\hat m,\mu) = \int d^3 x \,
  \left.\vev{ H(t,x) H(0)^\dagger}\right|_{g,\hat m,\mu}\, , 
\end{equation}
where we have indicated explicitly the dependence on the couplings and
the scale $\mu$. It is useful in this context to introduce a rescaled
mass $\hat m(\mu) = m(\mu)/\mu$.  For the specific case of lattice
simulations, the scale is set by the inverse lattice spacing
$\mu=a^{-1}$. The masses of the physical stable states are obtained
from the Euclidean time dependence of two-point functions. At large
Euclidean time $t$:
\begin{equation}
  \label{eq:expt}
  C_{H}(t;g,\hat m,\mu)\sim e^{-M_H t}\, , 
\end{equation}
where $M_H$ is the mass of the lightest state in the channel under
examination. We  examine the consequences of the
RG equation for the two-point function.

In the vicinity of the fixed point, a RG
transformation acts on the correlator according to:
\begin{equation}
  \label{eq:step1}
  \mu = b \mu' \;; \qquad C_{H}(t;g,\hat{m},\mu) = b^{-2
    \gamma_H} C_{H}(t;g^\prime,\hat{m}^\prime,\mu^\prime) \, ,
\end{equation} 
where $\gamma_H$ is the anomalous dimension of the field $H$.
The flow of the couplings near the RG fixed point is power-like:
\begin{equation}
g^\prime = b^{y_g} g\, ,~~~~ \hat m^\prime = b^{y_m} \hat m\, .
\end{equation}
We shall neglect henceforth the irrelevant coupling $g$ ($y_g < 0$). Multiplying
all mass units by the factor $b$ we obtain:
\begin{equation}
  \label{eq:step2}
  \qquad C_{H}(t;\hat{m}^\prime,\mu^\prime) = b^{-2 d_{H} }
  C_{H}(t b^{-1};\hat{m}^\prime,\mu) \;,
\end{equation} 
where $d_H$ is the naive mass dimension of the operator $H$. 
Choosing $b$ such that $\hat m^\prime=1$, the equations above yield:
\begin{equation}
  \label{eq:callansol2}
  C_{H}(t;\hat m,\mu) = \mathcal C_H F(t \hat m^{1/(1+\gamma_*)},\mu)\, ,
\end{equation}
where $F$ is some function that, for fixed $\mu$,  depends on the rescaled variable $x=t
\hat m^{1/(1+\gamma_*)}$ only. The detailed dependence of the 
prefactor $\mathcal C_H$ on the parameters
of the theory is postponed to the next section, where it will play a
prominent role. Comparing Eq.~(\ref{eq:callansol2}) with the expected
behaviour Eq.~(\ref{eq:expt}) yields:
\begin{equation}
  \label{eq:Mscale}
  M_H \simeq c_H \mu\,  \hat m^{\frac{1}{1+\gamma_*}}  
  \quad {\rm as} \quad m \to 0\, .
\end{equation}
Note that the scaling of the mass $M_H$ is entirely determined by the
anomalous dimension $\gamma_*$ and does not depend on the specific
choice of the interpolating operator $H$. \refeq{Mscale} shows that
all lowest state masses scale with with same exponent $1/(1+\gamma_*)$, while the
proportionality constant $c_H$ depends on the chosen channel. While
each individual mass in the spectrum vanishes, ratios of masses should
remain constant as the chiral limit is approached. This scaling is
consistent e.g. with the scenarios proposed in
Ref.~\cite{Miransky:1998dh,Luty:2008vs}.

In the derivation above we have not considered the effects of a finite
decay width. At least one channel ought to be stable and therefore not
affected by the width.  According to an inequality by
Weingarten~\cite{Weingarten:1983uj}, valid for $n_F \geq 2$, this
should be the mass of the lowest pseudoscalar flavour-nonsinglet,
which we shall later on denote by $M_{P^a}$.  For all other states one
might wonder how the width interferes with the derivation above. Could
the width and the mass conspire to cancel their leading mass scaling
behaviour in such a way as to invalidate Eq.~\eqref{eq:Mscale}?  We
would like to bring forward two reasons why this should not be the
case. First the difference in the large $N_c$-scaling of mass and
width ($\Gamma_H/M_H \sim {\cal O}(1/N_c)$) from QCD should hold in
mCGT too and serve as a parametric argument against such a
cancellation. Second we show in appendix~\ref{app:width} that in the
approximation where the self-energy is treated as being constant such
a cancellation can be excluded. This seems intuitively plausible since
in Euclidian time the mass and decay width behaviour are associated
with exponential and oscillatory behaviour respectively.

On the contrary since mass and width do not seem to interfere in the
leading large $t$-behaviour Eq.~\eqref{eq:callansol2} suggests that
both the mass and the width of the resonance scale according to
\begin{equation}
  \label{eq:doublescal}
  M, \Gamma \sim m^{1/(1+\gamma^*)}\, , 
\end{equation}   
We shall revisit the scaling of the width in Sect.~\ref{sec:fPS},
after discussing the scaling of the decay constants and derive
$\Gamma(A \to B + C) \sim m^{1/(1+\gamma^*)}$ for a specific decay
$A \to B + C$.

The behaviour \eqref{eq:Mscale} is markedly different from what is
observed in the spectrum of theories where chiral symmetry is
spontaneously broken, like {\it e.g.}\ in QCD. In the latter theories,
the Goldstone bosons become massless in the chiral limit, while the
other states remain massive, with their masses being of the order of
some typical hadronic scale $\Lambda$. For theories with an IRFP, all
states become massless, presumably at the same rate, which prevents  a
simple description of the nonperturbative low energy hadronic dynamics 
in terms of an effective theory like chiral perturbation theory.

Let us conclude this section by recalling how the finite-size effects
can be analyzed using RG equations. We shall discuss explicitly the case of the
correlator $C_H$, including the dependence on the size of the system
$L$.  We remind the reader that by studying finite volume effects, it
is implied that the box is larger than the typical scale,
$L \gg \mu^{-1}$, and therefore does not interfere with characteristic
short distance dynamics.  The solution of the RG equation, including
the $L$-dependence, scales as,
\begin{equation}
  \label{eq:callanFSS}
  C_H(t;\hat m,L,\mu) = b^{-2\gamma_H}
  C_H(t;\hat m^\prime,L,\mu^\prime)\, ,
\end{equation}
according to a modified version of Eq.~\eqref{eq:step1}.
Rescaling the energies by the factor $b$, and using
the power-law scaling of the couplings near the IRFP yields:
\begin{equation}
  \label{eq:FSSsol}
  C_H(t;\hat m,L,\mu) = b^{-2(d_H+\gamma_H)}
  C_H(b^{-1} t; b^{y_m} \hat m,b^{-1}L,\mu)\, .
\end{equation}   
Choosing $b$ such that $b^{-1} L=L_0$, where $L_0$ is a
reference length, yields:
\begin{equation}
  \label{eq:FSSscal}
  C_H(t;\hat m,L,\mu)=\left(\frac{L}{L_0}\right)^{-2\Delta_H}
  C_H\left(\frac{t}{L/L_0};x \frac{1}{\mu L_0^{y_m}},L_0,\mu\right)\, ,
\end{equation}
where we have introduced the scaling variable $x=L^{y_m}m$.

Comparing Eq.~\eqref{eq:FSSscal} with the expected asymptotic
behaviour in Eq.~\eqref{eq:expt} we obtain:
\begin{equation}
  \label{eq:MFSS}
  M_H = L^{-1} f(x)\, ,
\end{equation}
where $f(x)$ is some function of the scaling variable $x$, expected to vanish when
$x$ goes to zero. 
In order to recover the correct scaling with $m$ in the thermodynamic
limit
\begin{equation}
  f(x) \sim x^{1/y_m}\, ,\quad \mathrm{as}\ x\to\infty\, .
\end{equation}
As one can see from Eq.~\eqref{eq:MFSS}, if the fermion mass is
decreased at fixed $\mu$ and $L$, then the mass of the states in the
spectrum will initially decrease until the Compton wavelength of the
states is of the order of the linear size of the system. When this
happens, the mass of the states saturates and scales with the inverse
size $L^{-1}$. Results for $M_H L$ computed on different volumes
should follow a universal curve when studied as a function of the
scaling variable $x$.
\section{Modified Banks-Casher relation}
\label{sec:BCS}
In this section we relate the scaling exponent of the chiral
condensate $\eta_{\bar q q }$ to the scaling of the eigenvalue density
of the massless Dirac operator. We then illustrate how the
RG equations yield a prediction for the exponent in
terms of the anomalous dimension $\gamma_*$ introduced in
Eq.~\eqref{eq:gammadef}. These results follow readily from the RG
scaling of the free energy and the field correlators in the vicinity
of fixed point, and were already presented in
Ref.~\cite{DeGrand:2009et}. Here we discuss in detail the derivation
of these results in the context of a mCGT,  generalizing to other condensates such as the gluon condensate, and compare them to a more
heuristic derivation. 

\subsection{Eigenvalue density $\rho(\lambda)$ and the scaling exponent $\eta_{\bar q q}$}
\label{sec:eigenvalue}

It is useful to recall the basic steps in the derivation of the
Banks-Casher formula, in order to highlight the order in which the
limits are taken, the divergences that may appear, and 
 to
identify the differences from the case of a conformal theory.

We closely follow the discussion in Ref.~\cite{Leutwyler:1992yt} and
extend it at appropriate places to mCGT. The fermion propagator can be
written as:
\begin{equation}
  \label{eq:fprop}
  \langle q(x) \bar q(y) \rangle = \sum_n \frac{u_n(x)
    u_n^\dagger(y)}{m-i\lambda_n} \, ,
\end{equation}
where  the eigenmodes of the massless Euclidean
operator $D \equiv \gamma_\mu D^\mu$  have been introduced:
\begin{equation}
  \label{eq:Deigen} 
  D u_n(x) = \lambda_n u_n(x)\, .
\end{equation}
Since the eigenfunctions occur in pairs with opposite eigenvalues, the
chiral condensate in a finite volume $V$ is given by:
\begin{equation}
  \label{eq:cond1}
  \tchiral_V=\frac1V \int dx\, \vev{\bar q(x) q(x)} = - \frac{2m}{V}
  \sum_{\lambda_n > 0} \frac{1}{m^2+\lambda_n^2} \quad .
\end{equation}
Taking the infinite volume limit at fixed mass, the sum over positive
eigenvalues can be replaced by:
\begin{equation}
  \label{eq:cond2}
  \tchiral = \lim_{V\to\infty}\tchiral_V=-2m \int_0^\infty d\lambda
  \frac{\rho(\lambda)}{m^2+\lambda^2}\,,
\end{equation}
where $\rho(\lambda)$ denotes the number density of eigenvalues per
unit volume.  Eq.~\eqref{eq:cond2} is purely formal at this stage in the
sense that a UV-regularization is needed on both sides.
In four dimension the divergences are logarithmic and quadratic
respectively\footnote{Note that if the regulated theory breaks chiral
  symmetry explicitly, as is the case with lattice Wilson fermions,
  then a cubic divergence appears that survives in the chiral limit
  \cite{Hasenfratz:1998jp}.}.  The divergences can be isolated via a
twice-subtracted spectral representation:
\begin{eqnarray}
  \label{eq:cond3}
  \tchiral &=& -2m \int_0^\mu d\lambda
  \frac{\rho(\lambda)}{m^2+\lambda^2}   -2m^5 \int_\mu^\infty \frac{d\lambda}{\lambda^4}
  \frac{\rho(\lambda)}{m^2+\lambda^2} + \gamma_1 m + \gamma_2 m^3\, .
\end{eqnarray}
The subtraction constants $\gamma_1$ and $\gamma_2$ contain the 
UV-divergences.  Their respective behaviours are $\gamma_1 \sim
\Lambda_{\rm UV}^2$, and $\gamma_2 \sim\log \left[ \Lambda_{\rm UV}^2
  \right]$, and their actual values depend on two physical
renormalization conditions used to define the finite condensate on the
LHS of Eq.~\eqref{eq:cond2}. We shall investigate the limiting
behaviour when $m \to 0$.  
The second integral and the subtraction terms in Eq.~\eqref{eq:cond3} vanish
in the chiral limit ($m \to 0$).  Therefore only the first integral,
sensitive to the IR region, can result in a non-analytic term and has
to be investigated further.  A simple change of variable yields:
\begin{equation}
  \label{eq:cond4}
  \tchiral = -2 \int_0^{\mu/m} dx \,\frac{\rho(mx)}{1+x^2}  + {\cal A}(m) \, ,
\end{equation}
where ${\cal A}(m)$ stands for an analytic function of $m$.  From
Eq.~\eqref{eq:cond4}, following the same arguments used in QCD, one
can readily obtain:
\begin{equation}
  \label{eq:scaling}
  \vev{ \bar q q } \stackrel{m \to 0}{\;\; \sim \;\;} m^{\eta_{\bar
      qq}} \quad \Leftrightarrow \quad \rho(\lambda) \stackrel{\lambda
    \to 0}{\;\; \sim \;\;} \lambda^{\eta_{\bar qq}} \;.
\end{equation}
This in turn implies:
\begin{equation}
\eta_{\bar q q}|_{\rm QCD-like}  = 0 \;, \qquad  \eta_{\bar q q}|_{\rm mCGT}  > 0 \;,
\end{equation}
since in QCD the condensate remains finite in the chiral limit, while
it vanishes in mCGT.

Let us derive the same scaling coefficient $\eta_{\bar qq}$
\eqref{eq:BC3} from a RG analysis. The starting
point is the two-point function $C_{\bar qq}(t;\hat{m},\mu)$, as in
Eq.~\eqref{eq:twopt}, where the hadronic field $H = \bar qq$, and the
explicit dependence on the coupling $g$ is suppressed. The solution of
the RG equations for this specific case is:
\begin{equation}
  \label{eq:step2'}
  C_{\bar qq}(t;\hat{m},\mu) = b^{-2 \Delta_{\bar q q} } C_{\bar qq}(t
  b^{-1};b^{\ym} \hat{m},\mu) \;.
\end{equation}
Imposing again $b^{\ym}\hat{m} = 1$, finally leads to:
\begin{equation}
 C_{\bar qq}(t;\hat{m},\mu) = \hat m^{\frac{2 \Delta_{\bar
 qq} }{\ym}} C_{\bar qq}(t \hat m^{1/\ym} ;1,\mu) \;.
\end{equation}
Inserting a complete set of states the exponential decrease of any
state other than the vacuum for large $t$ results in:
\begin{equation}
  C_{\bar qq}(t;\hat{m},\mu) \stackrel{t \to \infty}{\sim} 
    m^{2 \eta_{\bar q q}}   \;,
\end{equation}
whence the scaling exponent~\eqref{eq:scaling} follows: 
\begin{equation}
  \label{eq:2}
  \eta_{\bar qq} = \frac{\Delta_{\bar q q}}{y_m} = \frac{3-\gamma_*}{1+\gamma_*} \;.
\end{equation}
The eigenvalue density then scales as:
\begin{equation}
\rho(\lambda) \sim \lambda^{(3-\gamma_*)/(1+\gamma_*)}\, ;
\end{equation}
this result generalizes the Banks-Casher relation for
QCD~\cite{Banks:1979yr}:
\begin{equation}
\left.\quark\right|_{m=0} \neq 0
\Rightarrow \rho(0) = - \pi \left.\quark\right|_{m=0}
\end{equation}
to mCGT. It is interesting to remark that Refs.~\cite{DeGrand:2009et}
and~\cite{Fodor:2008hm,Fodor:2009ar} state different predictions for the scaling
exponent. Our determination of this critical exponent agrees with
Ref.~\cite{DeGrand:2009et}.

Surely this derivation generalizes to any other operator, for example
the gluon condensate for which one gets:
\begin{equation}
\label{eq:G1}
\eta_{G^2} = \frac{\Delta_{G^2}}{y_m} = \frac{4}{1+\gamma_*} \;.
\end{equation}
The scaling dimension of the gluon condensate is four since it appears 
in the Lagrangian density of a four dimensional scale invariant theory.

\subsection{Alternative and heuristic derivation of $\eta_{\bar q q}$}
\label{sec:heuristic}

Let us now present an alternative derivation of the scaling exponents
$\eta_{\bar qq}$ and $\eta_{G^2}$ in Eqs.~\eqref{eq:2},~\eqref{eq:G1},
which is of a heuristic nature but might provide some physical
insight. The discussion for $\vev{\bar q q}$, which we shall adopt
here before generalizing it to $\vev{G^2}$ closely follows
Ref.~\cite{Sannino:2008nv}\footnote{The computation in
  Ref.~\cite{Sannino:2008nv} differs by in an additional term $\delta
  {\cal L} \sim (\bar q q)^2$ which is not relevant
  here.}\;\footnote{The calculation is similar to an analysis of a
  scale invariant theory with a scalar operator and tadpole term
  Ref.~\cite{Delgado:2007dx} in the context of the unparticle
  scenario, where $2 \leq \gamma_* \leq 1$ ($\Delta_{\cU} = 3 -
  \gamma_*$) was assumed and made it necessary to introduce (various)
  IR regularizations.}. In this work we refine the discussion and
interpretation of IR and UV-terms by making use of the scaling of the
hadronic masses in Eq.~\eqref{eq:Mscale} and the interpretation of
subtraction terms in Eq.~\eqref{eq:cond3}.

In a low energy effective theory describing the dynamics of the
operator $\bar q q\;$\footnote{We refrain to change to a notation
  $\bar q q \to {\cal O}_U$ since we are not interested in
  parametrizing an effective theory for ${\cal O}_U$ as in
  Ref.~\cite{Sannino:2008nv}.}, the mass deformation in
Eq.~\eqref{eq:mqq} corresponds to a tadpole term and demands a
reminimization of the potential to find the stable vacuum.  The
potential for $\bar q q$ is not known but the scaling of the two-point
function is governed by the anomalous dimension.  It has been proposed
in Ref.~\cite{Stephanov:2007ry} to mimic the continuous spectrum of
such an operator by introducing a tower of scalar fields with suitably
adjusted masses and couplings:
\begin{equation}
  \label{eq:dec}
  {\bar q q}(x) \sim \sum_n f_n \varphi_n(x) \, ; \qquad
  \matel{\varphi_n}{\bar q q}{0} \sim f_n \;, \qquad \left\{
\begin  {array}  {l}   \def  \ma  { \left(  \begin {array} {ccccc} } 
f_n^2 =   \delta^2  \,  (M_n^2)^{\Delta_{\bar q q} - 2} \\[0.1cm]
 M_n^2 = n \delta^2 \\
\end{array}  \right. \,,
\end{equation}
where the quantity $\delta$ describes the mass spacing between the
$\varphi_n$-modes. The decomposition \eqref{eq:dec} reproduces the
two-point function of a conformal theory in Minkowski space in the
limit $\delta \to 0$ \cite{Stephanov:2007ry}, up to potential
subtraction ambiguities.  Note that in Eq.~\eqref{eq:dec} we have not
tried to keep track of the overall mass dimension and normalization
since they are irrelevant for scaling properties.  The potential part
of the Lagrangian ${\cal L} = - m \sum_n f_n \varphi_n - 1/2 \sum_n
M_n^2 \varphi_n^2$ then leads to the equation of motion for
$\varphi_n$ of the form:
\begin{equation}
  \label{eq:phin}
  m f_n + M_n^2 \varphi_n =0 \quad \Rightarrow \quad \vev{\varphi_n} =
  -m f_n / M_n^2 \;,
\end{equation}
with solution as indicated on the right. Thus leading to a VEV,
\begin{alignat}{1}
\label{eq:condi}
& \vev{\bar q q} \sim  \sum_n f_n \vev{\varphi_n} = -m \sum_n \frac{f_n^2}{M_n^2} \;  \stackrel{\delta \to 0}{\rightarrow} \;   - m 
\int_{\Lambda_{\rm IR}^2}^{\Lambda_{\rm UV}^2} s^{\Delta_{\bar qq} - 3} ds \nonumber \\
& \qquad d(s) \; = \; \left\{ 
\begin  {array}  {ll}   \def  \ma  { \left(  \begin {array} {ccccc} } 
s^{\Delta_{\bar qq} - 3} &  \quad \Lambda_{\rm IR}^2 \leq s \leq \Lambda_{\cal U}^2   \\[0.1cm]
(\Lambda_{\cal U}^{2})^{\Delta_{\bar qq} - 3}  & \quad  \Lambda_{\cal U}^2   \leq s \leq \Lambda_{\rm UV}^2  \\
\end{array}  \right.
\end{alignat}
where IR- and UV-cutoffs, to be discussed below, were introduced.  We
have taken into account the running of the gauge coupling as indicated
in Fig.~\ref{fig:betafn}, though showing a somewhat cavalier attitude
towards the treatment of the transition region to be justified later
on.  The non-analytic part in $m$, if present, is hidden in the
IR-cutoff. It seems natural that the latter is governed by the typical
hadronic mass scale, i.e.  $\Lambda_\mathrm{IR} \simeq c M_H$, where
$c$ is a constant irrelevant to our investigations.  The integral
can be computed and yields:
\begin{equation}
  \label{eq:vac2}
  \vev{\bar q q} \sim -m \left(M_H^2\right)^{\Delta_{\bar q q}-2} + m
  (\Lambda_{\cal U}^2)^{\Delta_{\bar q q}-3} \Lambda_{\rm UV}^2 \, .
\end{equation}
Thus using the scaling of the hadronic masses \eqref{eq:Mscale} and \eqref{eq:Deltaym},
\refeq{vac2}  becomes
\begin{equation}
  \label{eq:scal2}
  \vev{\bar q q} \sim m^{\frac{3-\gamma_*}{1+\gamma_*}}\, + {\cal A}(m)
  \qquad \Rightarrow \qquad \eta_\qbarq = \frac{3-\gamma_*}{1+\gamma_*}\,
  ,
\end{equation}
where as previously ${\cal A}(m) \sim {\cal O}(m)$ denotes an analytic
function in $m$.  We have therefore derived the exponent
$\eta_{\qbarq}$ in \refeq{BC2}.
The UV-divergent term in Eq.~\eqref{eq:vac2} corresponds to the
quadratic divergence discussed in the previous section and is
irrelevant for the non-analytic part and the scaling exponent
$\eta_{\bar qq}$. The non-appearance of the logarithmic divergence
might be related to the fact that we do not consider the back reaction
of the mass perturbation on the spectrum such as taking into
consideration power correction in $m$ in the couplings $f_n$.

Surely this procedure generalizes to any gauge invariant term in the Largangian
$\delta {\cal L} = m^{(4-\Delta_{\cal O})/y_m} {\cal O}$ in which case the condensate \eqref{eq:scal2} assumes the form: 
\begin{equation}
\label{eq:genVEV}
\vev{{\cal O}}|_{\rm IR} \sim m^{\frac{4-\Delta_{\cal O}}{y_m}} (M_H^2)^{\Delta_{\cal O}-2}  \sim 
m^{\frac{\Delta_{\cal O}}{y_m}}  \;,
\end{equation}
in accordance with Eq.~\eqref{eq:2} which is absolutely general. The subscript IR indicates
that UV-terms have been omitted.

We consider it worthwhile to discuss the term $\delta{\cal L} \sim G^2$, resulting in
the gluon condensate. From Eq.~\eqref{eq:genVEV}, paying attention to the UV-terms in addition one gets:
\begin{equation}
\vev{G^2} \sim (M_H^2)^{\Delta_{G^2}-2} +  (\Lambda_{\rm UV}^2)^{\Delta_{G^2}-2}
\end{equation}
and with \eqref{eq:Mscale} and $\Delta_{G^2} = 4$ one gets :
\begin{equation}
\vev{G^2}  \sim m^{\frac{2(\Delta_{G^2}-2)}{y_m}} + {\cal A}(m^0) \quad  \Rightarrow  \quad \eta_{G^2} 
= \frac{2(\Delta_{G^2}-2)}{y_m}  = \frac{4}{1+\gamma_*}  \;,
\end{equation}
the same scaling as in Eq.~\eqref{eq:G1}. The ${\cal A}(m^0)$ refers to the term proportional to
$\Lambda_{\rm UV}^4$. It originates from the region of asymptotic freedom and is interpreted as a mixing 
with the identity. Such contributions, sometimes called renormalons, have hitherto prevented an
extraction or a proper definition of a gluon condensate in lattice QCD.

\subsection{Critical discussion}

Although the derivation above reproduces the correct result there
remain some points that deserve further clarification.
For this discussion we shall first think of the theory when $m = 0$.
Assuming that this effective field theory approach, the deconstruction \eqref{eq:dec},  can be extended to higher orders one would need to introduce higher order terms, starting from cubic ones, 
 into the effective Lagrangian in order to reproduce higher correlation functions.
There are two important effects, due to higher order terms.

First, are they going to modify the leading order extraction of the $\vev{\bar qq }$ directly?
The answer appears to be no, since the initial potential does not know about the
small parameter $m$ and perturbing the system by the term in
Eq.~\eqref{eq:mqq} does not lead to major correction to any order but
in the linear one. 
That this is self-consistent
can be seen more explicitly by plugging in the VEV $\vev{\varphi_n}
\sim m$ as in Eq.~\eqref{eq:phin} into a fictitious higher order
term.
Second, does the quadratic order need to be modified? 
Since the higher order terms are going 
to modify the two-point function, the answer appears to be yes. As long as these modifications are of the form
$M_n^2|_\text{higher order} \sim n^\alpha$ with $\alpha >1$, they will give rise
to subleading effects in $m$. Although this appears likely we have 
not tried to resolve this issue in this paper but obviously this question
deserves further study
\footnote{One might also be concerned that the mixing of different
$\varphi_n$ modes might reshuffle hierarchies. That this is not the
case for the analytic part follows from the fact that the modes with
low $n$ are responsible for the non-analytic part and that the higher
modes are numerically supressed w.r.t to the lower modes by
$\vev{\varphi_n} =
\vev{\varphi_1}/{n^{1+\gamma_*}}$~\cite{Sannino:2008nv}.}.
 
\subsection{Lattice data}
\label{sec:lateig}

Note that the anomalous dimension $\gamma_*$ is related to the running
of the fermion mass, and has been computed by Schr\"odinger Functional
methods in Ref.~\cite{Bursa:2009we}. On the other hand, as discussed
above, the exponent $\eta_\qbarq$ characterizes the behaviour of the
eigenvalue density around zero, so that in principle it can be
extracted from the eigenvalue density.

First results for the eigenvalue spectrum of the Dirac operator have
been presented in Refs.~\cite{DelDebbio:2008zf,DelDebbio:2010hu} for
an SU(2) gauge theory with two Dirac fermions in the adjoint
representation. An extensive study of the 200 lowest eigenvalues is
available only for the $16^3$ lattice studied in the references
above. As argued in Ref.~\cite{Giusti:2008vb}, the mode number of the
massive Dirac operator:
\begin{equation}
  \label{eq:modenumb}
  \nu(M,m) = \int_{-\Lambda}^{+\Lambda}d\lambda\, \rho(\lambda)\, ,
\end{equation}
where $\Lambda=\sqrt{M^2-m^2}$, carries the same information as the
density itself. The mode number can be renormalized and yields a
RG invariant, universal quantity that describes the
physical properties of the Dirac spectrum independent of the
regularization used.

\begin{figure}[ht]
  \centerline{\includegraphics[width=3.0in]{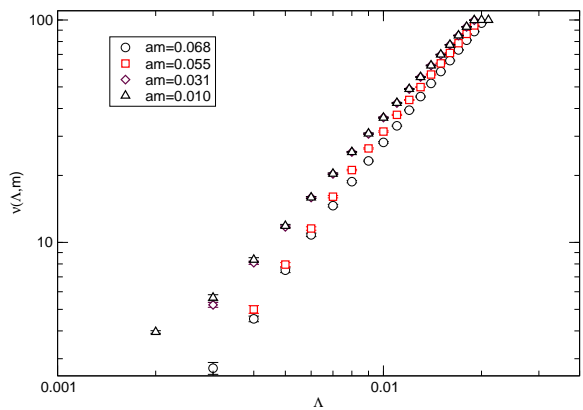}}
  \caption{\small The mode number $\nu(M,m)$ for the SU(2) gauge theory with
    two fermions in the adjoint representation. The data show the
    dependence of the mode number on the scale $\Lambda$ for several
    values of the quark mass. }
  \label{fig:modenum}
\end{figure}

The mode number can easily be computed from the available eigenvalue
distributions. Results obtained at $\beta=2.25$ on a $32\times 16^3$
lattice are reported in Fig.~\ref{fig:modenum} for different values of
the fermion mass $am$. 
The quantity of interest is the extrapolation of the mode number to
the thermodynamic and massless limit:
\begin{equation}
  \label{eq:extrapnu}
  \lim_{m\to 0} \lim_{V\to\infty} \nu(M,m)\, .
\end{equation}
Using the current data at a single value of the volume, where a reasonable amount 
of data is available Ref.~\cite{DelDebbio:2010hu},  the first
extrapolation cannot be performed. 
We defer
this analysis for further studies as larger lattices become available,
and concentrate instead on the extrapolation to the chiral
limit. The data in Fig.~\ref{fig:modenum} show that there is a
dependence of the mode number on the PCAC mass. The data at the two
lightest masses are compatible within the statistical errors, and we
shall take the data at the lightest mass for our analysis.  

In the chiral limit, the mode number is expected to scale as:
\begin{equation}
  \label{eq:scalxtrap}
  \nu(\Lambda,0)= C (\Lambda-g)^{\eta_{\bar qq}+1}\, ,
\end{equation}
where the possibility of a non-vanishing spectral gap $g$, as
suggested for instance in Ref.~\cite{Kovacs:2009zj}, is taken into
account in the functional form of the fitting function. we therefore
end up with fitting the data to three parameters, namely $C,g$, and
the exponent. The data are consistent with such a power-law behaviour,
however the value of the exponent depends critically on the range used
for the fit.
\begin{figure}[ht]

\centerline{\includegraphics[width=3.0in]{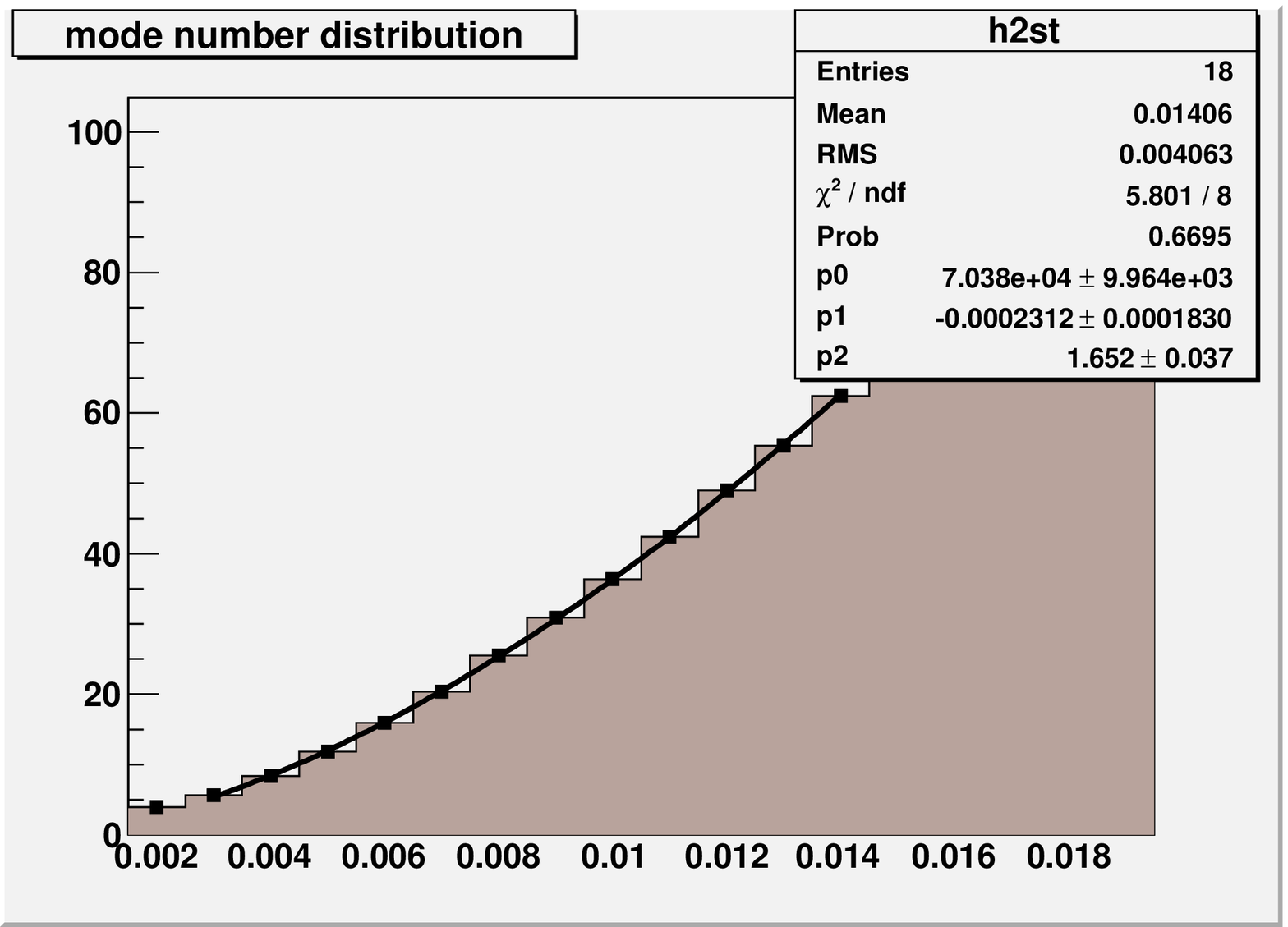},\includegraphics[width=3.0in]{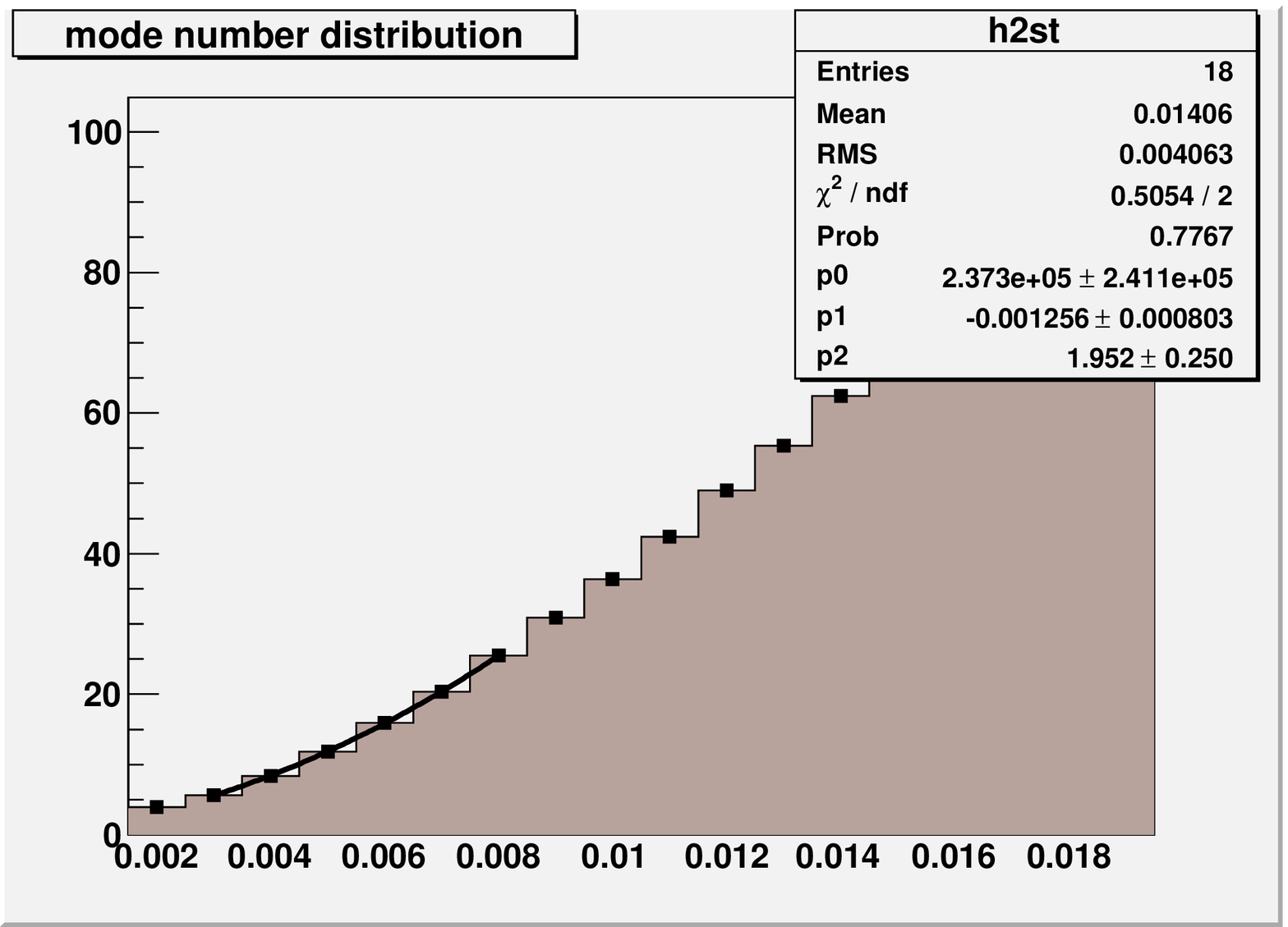}}
%
%
  \caption{\small Fit of the mode number $\nu(M,m)$ to a power law
  behaviour. The two plots represent two fits to the same data over
  two different ranges. 
  } \label{fig:fitmodenum}
\end{figure}
The plot in the left panel of Fig.~\ref{fig:fitmodenum} shows the
result of a fit that has a fit range $\Lambda\in [0.002,0.014]$, in lattice units. The result for
the critical exponent is $1+\eta_{\bar qq}=1.65(4)$, which can be
translated into a fitted value for the mass anomalous dimension,
yielding $\gamma_*=1.42(5)$. Note that this value is quite different
from the one found in previous
studies~\cite{DelDebbio:2010hu,Bursa:2009we}. However, if the fit
range is reduced to $\Lambda\in[0.003,0.008]$, then the fitted
exponent is $\gamma_*=1.1(3)$. The latter value is still larger
than the one obtained form the Schrodinger functional studies, it
clearly shows that the fitted exponent depends critically on the fit
range. More extensive data on the eigenvalue distributions are needed
in order to be able to extract the critical exponent in a reliable
manner. 

Note that the error induced by varying the fitting range turns out to
be larger than the statistical error, and that this result is obtained
at one value of the lattice volume, and could be affected by finite
size effects. A more comprehensive analysis of the volume
dependence of the eigenvalue distribution is needed. More extensive
lattice data should become available in the near future. 

\section{Decay constants}
\label{sec:fPS}
This section explores the information that can be gathered from the RG
scaling of matrix elements of given operators. In
subsection~\ref{sec:hyper_decay} we show that the scaling of decay
constants of the lowest lying states directly follow from its
anomalous dimension through the Callan-Symanzik equations.  In
subsection~\ref{sec:WI} we deduce consequences from spectral
representations of the Ward identities and low energy theorems for
pseudoscalar and scalar states evaluated at zero momentum.  In
subsection~\ref{sec:fPSdata} we compare our theoretical predictions
with recent lattice data. Finally miscellaneous matters of interest
are presented in section~\ref{sec:further}.  In appendix~\ref{sec:OPE}
the pseudoscalar WI at large momentum transfer is used to deduce
further information.

\subsection{Hyperscaling and decay constants}
\label{sec:hyper_decay}
Let us consider an operator ${\cal O}$ with scaling dimension
$\Delta_{\cal O}$ and quantum numbers such that it couples to a
state $|H(p)\rangle$ with strength $G_H$ for scalar operators and $F_H$ for vector operators. We shall choose $G_H$ to exemplify the equations below:
\begin{equation}
  \matel{0}{{\cal O}(0)}{H(p)} = G_H \;, \qquad \Delta_{\cal O} =
  d_{\cal O} + \gamma_{\cal O} \;.
\end{equation}
Information on the lowest lying state can be gained from the large
time behaviour of the Euclidian two-point function $C_{\cal
  O}(t;g,m,\mu)$ defined in Eq.~\eqref{eq:twopt}:
\begin{equation}
  C_{\cal O}(t;g,m,\mu) \stackrel{t \to \infty}{\longrightarrow}
  e^{-M_H t}\frac{\matel{0}{{\cal O}(0)}{H(p)}\matel{H(p)}{{\cal
  O}(0)}{0}}{2M_H V} = e^{-M_H t }\frac{|G_H|^2}{2M_H V}
\end{equation}
The scaling of $|G_H|$ can be inferred by applying a renormalization
group transformation $\mu = b \mu'$ and imposing $b^{\ym} \hat m(\mu)
= 1$ as in Sect.~\ref{sec:eigenvalue}. The LHS becomes:
\begin{equation}
\label{eq:LHS}
C_{\cal O}(t;\hat{m},\mu) = \hat m^{\frac{2
      \Delta_{\cal O}}{\ym}} C_{\cal O}(t \hat m^{1/\ym} ;1,\mu) \;,
\end{equation}
whereas the RHS scales as 
\begin{equation}
\label{eq:RHS}
\frac{|G_H|^2}{2M_H V} \sim \hat{m}^{2 \eta_{G_H} - 1/\ym + 3/\ym} \;.
\end{equation}
Combining Eqs.~\eqref{eq:LHS} and \eqref{eq:RHS} we obtain: 
\begin{equation}
\label{eq:resi}
|G_H| \sim \hat m^{ \frac{\Delta_{\cal O} - 1}{\ym}   } \;.
\end{equation}
The definitions of the decay constants, their anomalous dimensions and
resulting scaling coefficients are summarized in
Tab.~\ref{tab:tab_decay}.

We would like to the draw the reader's attention to the fact that the
pseudoscalar decay constant, as defined in Tab.~\ref{tab:tab_decay},
is related through the PCAC relation as
\begin{equation}
\label{eq:fg}
 \partial \Cdot A^a = 2m P^a   \;, \quad \Rightarrow \quad  2 m G_{P^a} = 
 M_{P^a}^2 F_{P^a} \;.
\end{equation}
The scaling is consistent with our findings from Eq.~\eqref{eq:resi} since:
\begin{equation}
  \label{eq:checkG}
  1+(2-\gamma_*)/\ym = 2/\ym + 1/\ym \, .
\end{equation}
Let us briefly discuss the scaling dimensions of the operators given
in Tab.~\ref{tab:tab_decay}.  The currents $V$, $V^a$ have vanishing
anomalous dimension since they are  conserved currents that are
associated with global symmetries. The axial current $A^a$ is only
partially conserved, see Eq.~\eqref{eq:fg}.  It is broken by a soft
term whose renormalization does not affect the divergence
$\partial \Cdot A^a$ and therefore $A^a$ has vanishing anomalous
dimension. Moreover this implies that $mP^a$ is a renormalization
group invariant and thus $\Delta_{P^a} = 3 -
\gamma_*$.  The scaling dimension of $S$ was already discussed in
section~\ref{sec:IRFP}.  In the case where there are no masses $S^a$
and $P^a$ have the same renormalization constant. This is explicit to
all orders in perturbation theory and should also hold
non-perturbatively. Neglecting effects of the mass on $\gamma_*$ one
concludes $\Delta_{S^a} = 3-\gamma_*$. The flavour singlet axial
vector identity is anomalous.  The topological charge density mixes with the
axial vector, which therefore does not renormalize multiplicatively.
This is further  discussed  in
appendix~\ref{sec:P}

\begin{table}[ht]
  \centering
  \begin{tabular}[h]{l | l | l | l | l | l}
    \renewcommand{\arraystretch}{1.4}\addtolength{\arraycolsep}{3pt}
    ${\cal O}$ & {\rm def} & $\matel{0}{ {\cal O}(0) }{ J^{\rm
    P(C)}(p) }$ & $J^{\rm P(C)}$ & $\Delta_{\cal O} = d_{\cal O}
    + \gamma_{\cal O}$ & $\eta_{G[F]}$ \\[0.1cm] \hline $S$ & $\bar q
    q$ & $G_{S}$ & $0^{++}$ & $3 - \gamma_*$ &
    $(2-\gamma_*)/\ym$ \\[0.1cm] $S^a$ & $\bar q \lambda^a q$ &
    $G_{S^a}$ & $0^{+}$ & $3 - \gamma_*$ &
    $(2-\gamma_*)/\ym$ \\[0.1cm] $P^a$ & $\bar q i\gamma_5 q$ &
    $G_{P^a}$ & $0^{-}$ & $3 - \gamma_*$ &
    $(2-\gamma_*)/\ym$ \\[0.1cm] $V$ & $\bar q \gamma_\mu q$ &
    $\epsilon_\mu(p) M_V F_{V} $ & $1^{--}$ & $3$ & $1/\ym$ \\[0.1cm]
    $V^a$ & $\bar q \gamma_\mu \lambda^a q$ & $\epsilon_\mu(p) M_V
    F_{V^a} $ & $1^{-}$ & $3 $ & $1/\ym$ \\[0.1cm] $A^a$ & $\bar
    q \gamma_\mu \gamma_5 \lambda^a q$ & $\epsilon_\mu(p) M_A
    F_{A^a} \,[i p_\mu F_{P^a}] $ & $1^{+}\,[0^{-}]$ & $3$ & $1/\ym\,
    [1/\ym]$ 
  \end{tabular} 
  \caption{\small Scaling laws, $G[F]\sim m^{\eta_{G[F]}}$, for the decay
    constants of the lowest lying states.  In regard to the $V/A$
    decay constants and formula \eqref{eq:resi} note that $G_{V/A} \leftrightarrow
    M_{V/A}F_{V/A}$ in terms of counting scaling powers. 
    The symbol $\ym \equiv 1 + \gamma_*$ denotes the
    scaling dimension of the mass \eqref{eq:Deltaym}. 
    Recall that the lowest bound state
    scales as $M_H \sim m^{1/\ym}$ \eqref{eq:Mscale} and the
    non-analytic part of the quark condensate is given by: 
    $\vev{\bar q q} \sim m^{(3-\gamma_*)/\ym}$ \eqref{eq:BC3}. 
     The symbol $a$ denotes the
    adjoint flavour index, and $\lambda^a$ are the generator
    normalized as ${\rm tr}[\lambda^a \lambda^b]
    =2\delta^{ab}$.}  
  \label{tab:tab_decay}
\end{table}

\subsection{Low energy theorems from Ward Identities and alike}
\label{sec:WI}
In the previous subsection we have inferred the scaling laws of the
decay constants of the lowest lying states from the anomalous
dimensions. Further information can be obtained by analyzing WIs.

In appendix~\ref{app:WI} we recall the derivation of two standard WIs,
Eqs.~\eqref{eq:wardPa} and~\eqref{eq:ward_crewther}, and a low energy
theorem, Eq.~\eqref{eq:qqqq}:
\begin{eqnarray}
\label{eq:wards}
& &  (2 m)^2 \Pi_{P_aP_b}(0)  =  -  2m  \delta_{ab}  \vev{\bar qq}  \nonumber \\
& &  (2 m)^2 \Pi_{P P }(0) -  \Pi_{\tilde Q \tilde Q}(0)  =  
  -  4m   \vev{\bar qq} \nonumber \\
& & \Pi_{SS}(0)  =  -  \frac{\partial}{\partial m} \vev{\bar qq}  \;,
\end{eqnarray}
where $\Pi_{XY}(q^2)$ is the time ordered two-point
function, c.f. \eqref{eq:time}. Information on the decay constants can be gained by
investigating the dispersion representation of the two-point
functions:
\begin{equation}
\label{eq:dispi}
  \Pi_{P_aP_b}(q^2) = \frac{1}{\pi} \int_{\rm cut} ds 
  \frac{{\rm Im}\left[ \Pi_{P_aP_b}(s) \right]}{s-q^2-i0}
  + c   + d \, q^2 + ...\;,
\end{equation}
where we have chosen $\Pi_{P_aP_b}$ as representative for
definiteness.  The symbols $c$ and $d$ denote subtraction constants
due to UV-divergences of which only $c$ is relevant since $q^2 = 0$ in
the equations above~\footnote{Note $c$ vanishes for $\Pi_{\tilde
Q \tilde Q}(0)$ since the latter vanishes to all orders in
perturbation because $\tilde Q$ can be written as a total
derivative.}. At $q^2 = 0$ Eq.~\eqref{eq:dispi} writes,
\begin{equation}
  \label{eq:dispi2} \Pi_{P_aP_b}(0) = \frac{1}{\pi} \int_{\rm cut}
  ds \frac{{\rm Im}\left[ \Pi_{P_aP_b}(s) \right]}{s-i0} +c_1 + c_2
  m^2 \;,
\end{equation}
where $c = c_1 + c_2 m^2$ are the subtraction constants due to a
quadratic and logarithmic divergence of which only $c_1$ is relevant
for our discussion. 

We would like to make a comment of speculative nature. 
Assuming that the lowest state,
\begin{equation}
  \label{eq:WIdisp} 
  \Pi_{P_aP_b} (0)
  = \delta_{ab} \frac{G_{P^a}^2}{M_{P^a}^2} + \ldots \, ,
\end{equation}
contributes to the leading scaling of the RHS of Eqs.~\eqref{eq:wards}\begin{equation}
 \label{eq:gGMOR}
    \frac{G_{P^a}^2}{M_{P^a}^2}   +  \dots   + c_1 
  = -\frac{2}{m} \vev{\bar qq}\, ,
\end{equation}
then, using the results for $M_H$ and setting aside 
the issue of the subtraction constant $c_1 \sim {\cal O}(m^0)$  for the moment, 
the scaling laws in Tab.~\ref{tab:tab_decay} are reproduced for $P^a$ and $S$ and
\begin{equation}
 \frac{G_{P}^2
    - (\tilde G_{P}/2m)^2}{M_P^2} \sim m^{\eta_{\bar q q} - 1} =
    m^{\frac{2(1-\gamma_*)}{1+\gamma_*}} \, ,
\end{equation}
would follow from the pseudoscalar WI, where the decay constants $G_P$ and $\tilde G_P$ 
are defined as follows.
\begin{equation}
  \matel{0}{P(0)}{P(p)} = G_{P} \;, 
  \qquad \matel{0}{\tilde Q(0)}{P(p)} = \tilde G_{P}\, .
\end{equation}
The result for $P$, if correct, is new but on the
otherhand not very practical as it is the difference of two positive
terms.
Another possibility, or rather an extension of the scenario above, is that the hadron mass scaling is universal and that the width does not upset the parametric effects\footnote{This should be true in the large $N_c$-limit where the width is suppressed  by $1/N_c$ as compared to the mass.},
which would result in all decay constants scaling in the same way. Clearly these statements above are of a speculative nature driven by the knowledge of the lowest lying decay constants
from the Callan-Symmazik equations.
It is also
amusing to see in what way the scaling laws for those two channels
turn out to be the same. 
One might wonder about the influence and the origin of the subtraction
terms in the WI~\eqref{eq:wards}.  They match the divergences of the
quark condensate on dimensional grounds.  In fact, 
it can be seen that they match exactly.  Following Ref.~\cite{Sannino:2008nv} we can fix
the coupling 
\begin{equation}
\label{eq:match3}
\matel{0} {\bar q q}{\varphi_n} = \sqrt{
  \frac{B_{\Delta_{\bar q q}}}{2 \pi}} f_n
\end{equation}  
\eqref{eq:dec} at $\Delta_{\bar qq} = 3$ by
demanding that the deconstructed version matches $\Pi(0)_{SS}$ in the
region where the theory is asymptotically free. At vanishing quark mass 
and $O(g^0)$ the time dependent scalar correlator,  $\Pi(q^2)_{SS}$, is given by:\begin{equation}
  \Pi(q^2)_{SS} = \frac{B_3}{2\pi} \int \frac{s ds}{s-q^2} \;, \qquad
  B_3 = \frac{N_c n_f}{4\pi}\, .
\end{equation}
We have factored out the coefficient $B_3$, which matches the
deconstructed version.  The condensate~\eqref{eq:condi} in the
asymptotically free region, with the normalization \eqref{eq:match3}, becomes:
\begin{equation}
  \vev{\bar q q } =  - m \frac{B_3}{2\pi}  \int ds \Cdot 1\,.
\end{equation}
Whence an exact matching of the divergences in the correlation
function~\eqref{eq:wards} on the LHS, and the quark condensate on the
RHS is found. Thus the UV-divergences of the quark condensate and the ambiguity in 
defining the $\Pi(q^2)$ functions from a dispersion relation do match exactly. The deconstructed version \eqref{eq:dec} of the condensate is therefore consistent with the WI.
Note the extra factor of one half on the RHS of the
non-singlet pseudoscalar WI is due to the normalization ${\rm
tr}[\lambda^a \lambda^b] = 2\delta^{ab}$ of the flavour
generators\footnote{We have only focused on the leading quadratic
divergence of the quark condensate; it would be interesting to
investigate the logarithmic divergence as well in which case one could
possibly learn something about mass correction to the deconstructed
version presented in section \ref{sec:heuristic}.}.

\subsection{Comparison with data}
\label{sec:fPSdata}

\begin{figure}[ht]
  \centerline{\includegraphics[width=4.0in]{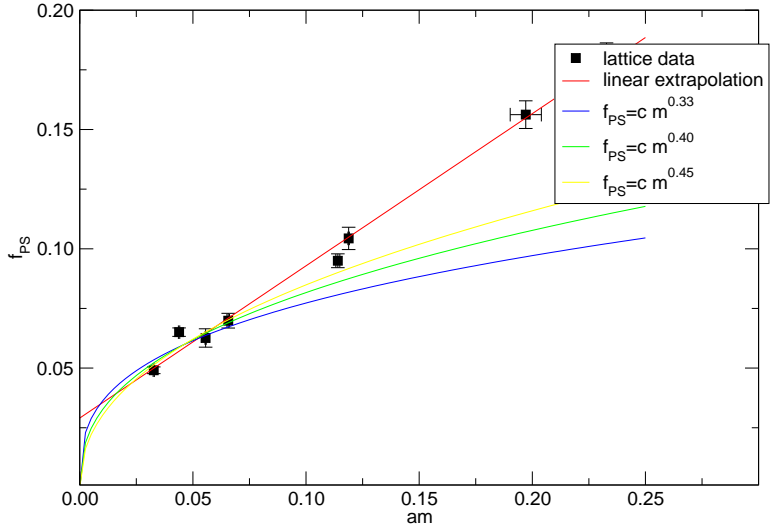}}
  \caption{\small The pseudoscalar decay constant $F_{P_a}$ as a function of the
    fermion PCAC mass $am$. All quantities are expressed in units of
    the lattice spacing, which is kept constant as $m$ is varied. The
    lines represent fits of the data to power--law scaling for
    different values of the critical exponents $\eta_{F_{P_a}}$.}
  \label{fig:FPS}
\end{figure}

The scaling predictions obtained above can be compared to the recent
lattice data presented in Ref.~\cite{Pica:2009hc} for the SU(2) gauge
theory with two flavors in the adjoint representation. The dependence
of the pseudoscalar decay constant on the fermion mass is reported in
Fig.~\ref{fig:FPS}. It is clear from the plot that the non-analytic
dependence of $F_{P^a}$ on the fermion mass can not be determined from
current lattice data, where no curvature is visible. This is confirmed
quantitatively by trying to fit the data to a power law dependence on
the fermion mass. Rather than trying to determine the exponent from
the fit, we keep the exponent fixed, and fit the proportionality
coefficient only. As shown in the figure, good fits to the data at the
smaller masses can be obtained for different values of
$\eta_{F_{P_a}}$ by adjusting the constant of proportionality $c$.  

A preliminary estimate for the mass anomalous dimension was obtained
from numerical simulations using the Schr\"odinger functional in
Ref.~\cite{Bursa:2009we}.  A value of $\gamma_* \simeq 0.5$ is
compatible with the results in Ref.~\cite{Bursa:2009we}, and leads to
$\eta_{G_{P_a}} \simeq 1.0$ and $\eta_{F_{P_a}} \simeq 0.33$. This is
represented by the blue line in the figure. As discussed above, the
value $c$ can be adjusted to yield a good description of the data at
the smaller masses. Note that we only expect the scaling to hold in
the limit where the fermion mass goes to zero, therefore it seems
natural to exclude the heavier points from this analysis. However
further systematic uncertainty is introduced by the choice of the
fitting range. 

\begin{figure}[ht]
\centerline{\includegraphics[width=4.0in]{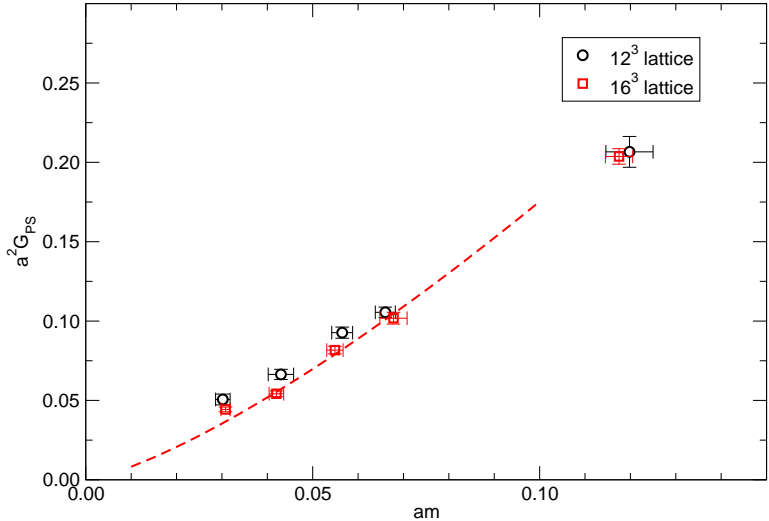}}
  \caption{\small The pseudoscalar coupling $G_{P_a}$ as a function of the
    fermion PCAC mass $am$. All quantities are expressed in units of
    the lattice spacing, which is kept constant as $m$ is varied. The
    line represents a fit to the data assuming the power-law behaviour
    described above.}
  \label{fig:GPS}
\end{figure}

The situation improves only slightly when looking at the dependence of
the coupling $G_{P^a}$ on the fermion mass, which is presented in
Fig.~\ref{fig:GPS}. A two-variable fit of the data can be performed in
this case, and yields a scaling exponent $\eta_{G_{P^a}}=1.3(2)$. Note that,
using the scaling formula in Tab.~\ref{tab:tab_decay}, the result of
the fit implies $\gamma_*=0.30(5)$, which broadly agrees with the
result of Refs.~\cite{Bursa:2009we,DelDebbio:2010hu,DelDebbio:2010hx}.

Better control over systematic errors is required in order to extract
robust information from the scaling of the pseudoscalar decay
constant. Current data can only be used to check the consistency with
the scaling we presented above; Fig.~\ref{fig:FPS} clearly shows that
lighter masses are needed in order to actually determine the scaling
exponents from lattice data.

\subsection{Further remarks}
\label{sec:further}

\subsubsection{Scaling of the decay width}
\label{sec:scalingwidthexample}

Let us focus on a generic decay process $A \to B C$, mediated by some
effective Lagrangian:
\begin{equation}
  \label{eq:Leff}
  \mathcal L_\mathrm{eff} = \int d^4x \, G_{ABC} A(x) B(x) C(x)\, ,
\end{equation}
where $G_{ABC}$ is the $ABC$ coupling, and $A(x),B(x),C(x)$ are the
fields creating and annihilating the states $A,B,C$. The fields are
normalized as $\langle 0|A(0)|A(p)\rangle = 1$ etc.  Let us now
introduce three interpolating fields $J_A$, $J_B$, and $J_C$; these
are composite fields that have an overlap with the single particle
states,  e.g. like  quark bilinears for the simplest type of mesons.  
At lowest order in
$G_{ABC}$ the correlator is given by:
\begin{eqnarray}
  \langle J_A(x) J_B(y) J_C(0)\rangle & \sim & \int d^4z \,
 G_{ABC} \langle J_A(x) J_B(y) J_C(0) \mathcal
 L_\mathrm{eff}(z) \rangle \;,
 \end{eqnarray}
 and inserting a complete set of states becomes,
 \begin{eqnarray}
  \label{eq:threecorr}
 &\sim& \frac{G_A G_B G_C}{M_A M_B
 M_C}\, \left[\frac{VT}{V^3}\right]\, G_{ABC} e^{- M_A(t_x-t_z)
 -M_B(t_y-t_z)-M_C(-t_z) }\quad .
\end{eqnarray}
Using the scaling laws discussed above, we find for the LHS of
Eq.~\eqref{eq:threecorr}:
\begin{equation}
  \label{eq:lhsscal}
  \langle J_A J_B J_C \rangle \sim
  m^{(\Delta_A+\Delta_B+\Delta_C)/y_m}\, ,
\end{equation}
while from the RHS we obtain:
\begin{equation}
  \label{eq:rhsscal}
  \langle J_A J_B J_C \rangle \sim
  m^{(\Delta_A-1)/y_m}\, m^{(\Delta_B-1)/y_m}\, m^{(\Delta_C-1)/y_m}\, 
  m^{-3/y_m}\, m^{5/y_m}\, G_{ABC}\, ,  
\end{equation}
and therefore
\begin{equation}
  \label{eq:gabscal}
  G_{ABC}\sim m^{1/y_m}\, .
\end{equation}
The scaling of $G_{ABC}$ determines the scaling of the decay width for
this specific channel:
\begin{equation}
  \label{eq:gamscal}
  \Gamma(A \to B + C) \sim \frac{\left|G_{ABC}\right|^2}{M_A} \sim m^{1/y_m}\, ,
\end{equation}
which corresponds to the same scaling we have argued for in section \ref{sec:IRFP} Eq.~\eqref{eq:doublescal}.

\subsubsection{Heavy quarks or mass deformed conformal?}

It has been pointed out \cite{DelDebbio:2009fd} that a confining theory with chiral symmetry
breaking, large mass term and small volume could mimic a conformal
theory with mass perturbation. Thus the question: how to distinguish a
conformal theory with small mass perturbation from a heavy quark
regime? We would like to emphasize that the scaling laws of the
pseudoscalar decay constant are a major help in this respect.  The
decay constant of pseudoscalar meson of two heavy quarks, which we
denote by $\bar b b$ in analogy with QCD, are expected to have the
following scaling behaviour:
\begin{equation}
  \label{eq:HQscal} 
  F_{\bar bb} \sim m^{-1/2} \;, \qquad ( \Rightarrow
  G_{\bar bb} \sim m^{1/2}) \; .
\end{equation}
This follows from the two heavy-quark state behaving like a quantum
mechanical bound state, and therefore being treated using a quark
model; Eq.~\eqref{eq:HQscal} expresses a result found a long time ago
by van Royen and Weisskopf~\cite{VanRoyen:1967nq}. This is clearly
different from the behaviour of the decay constant of a pseudoscalar
state, or any other state, in a mCGT, which vanishes in the limit
$m \to 0$, c.f. Tab.~\ref{tab:tab_decay}.

\section{Summary and conclusions}
\label{sec:concl}

In this paper we have accumulated a number of analytical results for 
mass scaling exponents $\eta$ of the type \eqref{eq:Om} for lowest state observables
from Callan-Symmanzik equations, which should help to identify the conformal window 
of four dimensional non-supersymmetric  gauge theories. Possibly the clearest evidence of the conformal
window so far comes from the vanishing of the hadron masses and decay constants 
in the chiral limit. We have identified the scaling of $F_{\bar bb} \sim m^{-1/2}$ as a criterion 
to distinguish mCGT from a heavy quark regime at small volume which has been pointed out
as a potential pitfall in identifying the conformal window \cite{DelDebbio:2009fd}.

In section \ref{sec:eigenvalue} mass scaling coefficients for condensates are 
determined from Callan-Symmanzik equations of which we
mention $\eta_{\bar q q} = (3-\gamma_*)/(1+\gamma_*)$ and
$\eta_{G^2} = 4/(1+\gamma_*)$. In section \ref{sec:heuristic} 
we provide a more physical, but somewhat more heuristic, derivations of those results
and discuss the nature of IR- and UV-regularizations within this framework.
By generalizing the Banks-Casher relation from QCD it is shown, in section \ref{sec:eigenvalue}, that the exponent of the eigenvalue density of
the Dirac operator is $\eta_{\bar qq}$, providing an alternative method for extracting $\gamma_*$. As our discussion in section \ref{sec:lateig} shows, 
it is too early to draw conclusions on the extraction of $\gamma_*$ by this method. 
In section \ref{sec:fPS} we derived scaling laws for all lowest state 
decay constants, summarized in Tab.~\ref{tab:tab_decay},
other than those affected by the chiral anomaly and of tensorial structure. Fitting to current data for
$F_{P_a}$ and $G_{P_a}$ we find results for $\gamma_*$ compatible with earlier derivations of $\gamma_* \simeq 0.5$.
Summarizing, the derivations  indicate that  lowest state observables ${\cal O}$  scale as:
\begin{equation}
\eta_{\cal O} = \frac{\Delta_{\cal O}}{1+\gamma_*} \;,
\end{equation} 
where the one-particle state, $|H(p)\rangle$, scaling dimension turns out to be
$\eta_{|H(p)\rangle} =  -1/(1+\gamma_*)$.
Whether or not this is true for higher states as well, remains unclear and is more of academic interest as higher states are difficult to assess on the lattice. The real distinction of higher states is presumably the decay width and the associated continuum thresholds. It is therefore tempting to think that in the large $N_c$-limit, where the width is supposed to vanish, all decay constants and masses in a specific channel scale in the same way as the lowest one.

To this end we would like to add a few comments on tentative conclusion on the
scaling of the $S$-parameter.
The electroweak $S$-parameter is  proportional to the $V$-$A$ correlator
evaluated at zero momentum, $\Pi_{V \text{-}A}(0) \sim \int   \rho_{V\text{-}A}(s)/s ds$, 
with the pion pole subtracted. A lattice computation of the $S$-parameter for 
walkinig technicolour (WTC) theories would be phenomenologically important. Thus
it is crucial to distinguish, in a parametric way, the regimes of WTC and mCGT\footnote{Of course this does not solve the problem of distinguishing the WTC from TC regime per se.}.
Assuming  that the correlator is saturated by the lowest resonances\footnote{This is reasonably satisfied in QCD. Adding one more triplets of states $P,V,A$ would not really alter the conclusions above.}
$\Pi_{V \text{-}A}(0)  \sim F_V^2/M_V^2 - F_A^2/M_A^2 - F_\pi^2/M_\pi^2$,
it is just the pion pole that serves as an indicator since 
$(M_\pi^2)_{WTC} \sim {\cal O}(m)$ leads to 
$\Pi^{\rm WTC}_{V \text{-}A}(0) \sim  {\cal O}(m^{-1})$  and the results in  Tab.~\ref{tab:tab_decay} imply   $\Pi^{\rm mCGT} _{V \text{-}A}(0) \sim  {\cal O}(1)$  .
To this end let us note that, since the pion mass is the lowest one in the 
spectrum \cite{Weingarten:1983uj}, a determination of $F_V^2/F_\pi^2 \leq 1$ would imply 
$\Pi^{\rm mCGT} _{V \text{-}A}(0) < 0$.

\section*{Acknowledgments}
LDD \& RZ gratefully acknowledge the support of advanced STFC fellowships. 
We are grateful to Biagio Lucini, Michela Petrini and Laurent Lellouch for discussions. We would also like to thank Biagio Lucini, Agostino Patella, Claudio Pica, and Antonio Rago for granting us access to the lattice data used in this study, and for several useful discussions. 

\paragraph{Note added:} 
We would like to add that in our more recent work \cite{DelDebbio:2010jy} we have established, invoking similar RG arguments, that the scaling relation for decay constants and masses are valid for the entire spectrum.
Furthermore, using the trace anomaly and the Feynman-Hellmann theorem we have derived the mass
scaling relations without resorting to RG arguments.

\appendix
\setcounter{equation}{0}
\renewcommand{\theequation}{A.\arabic{equation}}

\section{Operator product expansion in the deep Euclidian}
\label{sec:OPE}
In Ref.~\cite{Shifman:1978by} an OPE relation is obtained which we
shall derive here in a slightly modified form.  Taking the Fourier
transform of Eq.~ \eqref{eq:ward_start} one arrives at the expression
\begin{equation}
  2 m q^\mu  \int e^{i q \,\Cdot \, x} \vev{T A_\mu^a(x) P^b(0)}_0  =  
  (2m)^2\Pi_{P_aP_b}(q^2) + 2 m \vev{ \bar q q } \;.
\end{equation}
Inserting a complete set of states on the LHS and using
Eq.~\eqref{eq:fg} yields:
\begin{equation}
  \label{eq:OPE} 
  - \sum_P \frac{F_{\rm PS}^2 M_{\rm PS}^2}{M_{\rm
  PS}^2 + Q^2} + d_1 + d_2 Q^2 = \frac{(2
  m)^2}{Q^2} \Pi_{P_aP_b}(-Q^2)+ \frac{2 m \vev{ \bar q q }}{Q^2} \;,
\end{equation}
where $Q^2 \equiv -q^2$, and $d_1$, $d_2$ are subtraction constants,
which can also depend on $m$. Neither of these are of relevance to us
since we may simply differentiate this expression twice with respect
to $Q^2$. Following Ref.~\cite{Shifman:1978by}, an expansion in one
inverse power of $Q^2$ yields Eq.~\eqref{eq:gGMOR}, by assuming that
$\Pi_{P_aP_b}(-Q^2)$ does not vanish as $Q^2 \to \infty$ and $m \to
0$.  By expanding in one more inverse power in $Q^2$ one observes that
the scaling of $F_{\rm PS}^2 M_{\rm PS}^4$ has to be larger than $m^2$,
\begin{equation}
  \label{eq:a3}
  2 \eta_{F_{\rm PS}} + 4 \eta_{M_{\rm PS}}  \geq 2  \;.
\end{equation} 
In the domain where $1 \leq \gamma_*$, Eq.~\eqref{eq:a3} leads to
$\gamma_* \leq 2$, which corresponds exactly to the unitarity bound
$\Delta_{\bar q q} \equiv 3 - \gamma \geq 1$ for a scalar
field~\cite{Mack:1975je}.

\section{Ward Identities and a low energy theorem}
\label{app:WI}
In this appendix sketch the derivation of two standard WI and one low energy theorem. 

\subsection{Ward identity for  pseudoscalar non-flavor-singlet}
\label{eq:Pa}
Starting with the identity,
\begin{equation}
  \label{eq:ward_start}
  \partial^\mu \vev{0| T A_\mu^a(x)  P^b(0)|0 } = 
  \vev{0|T   \partial \Cdot A^a(x)  P^b(0)|0} +
  \delta(x_0)  \vev{ 0|  [A_0^a(x), P^b(0)]|0} \;,
\end{equation}
and integrating the equation over $d^4x$ one arrives at the
pseudoscalar Ward identity (WI)\footnote{We are using the fact that there
  are no Goldstone bosons since $\mathrm{SU}(n_f)_A$ is explicitly
  broken.}:
\begin{equation}
  \label{eq:wardPa}
  (2 m)^2 \Pi_{P_aP_b}(0)  =  -  2m  \delta_{ab}  \vev{\bar qq}\;,
\end{equation}
where 
\begin{equation}
\label{eq:time}
  \Pi_{P_a P_b } (q) =  i \int_x e^{iq \cdot x}  
  \matel{0}{T   P_a(x) P_b(0)}{0}\, . 
\end{equation}
Throughout the paper
$\Pi_{AB}(q)$  denotes the time-ordered two-point correlator of 
operators $A$ and $B$.  
The RHS of Eq.~\eqref{eq:wardPa} is obtained by evaluating the
commutator
\begin{eqnarray}
   \matel{0}{[Q_5^a|_{x_0=0}, \partial \Cdot A^b(0) ]}{0}  = 
  2 i  \delta^{ab} m \vev{\bar qq} \; .
\end{eqnarray}
The charge is defined as usual: $Q_5^a|_{x_0=0} = \int d^3 x
A^b(0,\vec{x}) $. 

\subsection{(Anomalous) Ward identity for pseudoscalar flavor-singlet}\label{sec:P}

The flavor singlet sector can be analysed by reviewing the Ward
identities \cite{Crewther:1977ce} used in discussing the
$\eta'$-mass/U(1)$_A$-problem in QCD.  Let us define the following
flavor singlet quantities:
\begin{eqnarray}
\label{eq:Qt}
  P(x) = \sum_{j=1}^{n_f} \bar q_j i \gamma_5 q_j(x)  \;, \quad
  A_\mu(x)  = \sum_{j=1}^{n_f} \bar q_j\gamma_\mu  \gamma_5 q_j(x) \;, \quad
  \tilde Q(x) = \frac{g^2}{16 \pi^2} n_f  
  \widetilde G_{\alpha \beta} G^{\alpha \beta}(x) \;,
\end{eqnarray}
where $\widetilde G_{\alpha \beta} = 1/2 \epsilon_{\alpha \beta \gamma
  \delta} G^{\gamma \delta}$.  The anomaly equation is given by:
\begin{equation}
\label{eq:anomaly}
 \partial \Cdot A = 2m P + \tilde Q
\end{equation}
The integrated anomalous Ward identity is readily obtained from
Eq.~\eqref{eq:ward_start} and reads:
\begin{equation}
  \label{eq:ward_P}
  (2 m)^2 \Pi_{P P }(0) + (2m) \Pi_{ \tilde Q P}(0)  =  -  4m   \vev{\bar qq} \;.
\end{equation}
By observing that 
\begin{equation}
  \label{eq:transition}
  0 = i \int_x   \partial_\mu \matel{0}{T \tilde Q(0) A^\mu(-x)}{0} = 
  (2m) \Pi_{\tilde Q P}(0)  +  \Pi_{\tilde Q \tilde Q}(0)  \;,
\end{equation}
the WI \eqref{eq:ward_P} can be written in a more symmetric form:
\begin{equation}
  \label{eq:ward_crewther}
  (2 m)^2 \Pi_{P P }(0) -  \Pi_{\tilde Q \tilde Q}(0)  =  
  -  4m   \vev{\bar qq} \;.
\end{equation}

\subsection{Low energy theorem for scalar flavour-singlet}
\label{sec:S}

A simple and usefel relation follows from  the fact that the operator $S(x) = \bar q q(x)$ appears in the Lagrangian.
Eq.~\eqref{eq:mqq} implies:
\begin{equation}
  \label{eq:qqqq}
  \Pi_{SS}(0)  =  -  \frac{\partial}{\partial m} \vev{\bar qq}  \;,
\end{equation} 
where $\Pi_{SS}(q^2)$ is the time ordered two-point function \eqref{eq:time}.

\section{The decay width}
\label{app:width}
In this appendix we investigate whether the large $t M_H$ behaviour of
a correlation function can be influenced by the width. 
According to \cite{Michael:1989mf} the Euclidian time behaviour of
a two-point function in the rest frame of the decaying 
particle can be written as a spectral integral of the type
\begin{equation}
  \label{eq:resdisp}
  C_H(t;g,m,\mu) = \frac{1}{\pi} \int dE e^{-Et} \rho(E)\, ,
\end{equation}
where 
\begin{equation}
  \label{eq:rhospect}
  \rho(E) = \frac{{\rm Im}(\Sigma(E))}{|m^2 - E^2 - \Sigma(E)^2|} 
\end{equation}
and $\Sigma(E)$ is the self energy and $m$ the bare mass.
We shall work in the approximation where $\Sigma(E)$ does not vary appreciably 
around the peak $E =  |M|$ and we neglect the far away singularity $E = -|M|$.
The symbol $M$ denotes the renormalized mass: 
$M =  m^2 - {\rm Re}(\Sigma(M))$. 
In this case $\rho(E)$ assumes the form:
\begin{equation}
\rho(E) = \frac{\gamma(E)}{(M-E)^2 + \gamma(E)^2} \;,
\end{equation}
where $\gamma = \Gamma/2 = {\rm Im}(\Sigma(E))/2M$. Then 
the two-point function takes the following form \cite{Michael:1989mf}:
\begin{equation}
C_H(t;g,m,\mu)  = \frac{e^{-Mt}}{2M} {\rm Ei}(\gamma t,(M-\lambda)t) \;,
\end{equation}
where $\lambda$ is the onset of the cut, omitted in Eq.~\eqref{eq:resdisp}, and
\begin{equation}
{\rm Ei}(\alpha,\beta) = \int_{-\beta}^\infty \frac{\alpha e^{-x}}{x^2 + \alpha^2}   dx  
> 0\;.
\end{equation}
If there are cancellations  between the $\sim e^{-Mt}$ 
behaviour of the mass and the width then the following must be true:
\begin{equation}
\label{eq:hypo}
{\rm Ei}(\alpha,\beta) \stackrel{t \gg M}{\sim} \alpha^n  e^{|b|\alpha} \;,
\end{equation}
where $n$ a real number and $b$ is a number that would have to be fined tuned.
It can be shown that this cannot be the case. Consider 
\begin{equation}
\alpha {\rm Ei}(\alpha,\beta) = \int_{-\beta}^\infty \frac{\alpha^2 e^{-x}}{x^2 + \alpha^2}   dx
\leq \int_{-\beta}^\infty e^{-x}   dx < \infty \;,
\end{equation}
but then from  \eqref{eq:hypo}:
\begin{equation}
\infty >  \alpha {\rm Ei}(\alpha,\beta)  \stackrel{t \gg M}{\sim}  \alpha^{n+1}  e^{|b|\alpha} \stackrel{\alpha \to \infty}{\to} 
\text{divergent}
\end{equation}
one gets an immediate contradiction to the hypothesis \eqref{eq:hypo}. Thus we have shown that,
in the approximations mentioned above, that a the large Euclidian time behaviour
of the mass and the width do not conspire to cancel each other.


\end{document}